\documentclass{sig-alternate-10pt}
\usepackage[margin=1in]{geometry}
\usepackage{microtype}
\usepackage{hyperref}
\usepackage{rotating}
\usepackage{amssymb}
\usepackage{color,soul}
\usepackage{adjustbox}
\usepackage{cite}
\usepackage{float}
\usepackage{subcaption}
\usepackage{balance}
\usepackage{pgfplots}
\pgfplotsset{compat=1.18}
\usepackage{numprint}
\usepackage{xspace}
\usepackage{comment}
\usepackage{xurl}
\usepackage{tabularx,ragged2e}
\usepackage{tikz}

\usepackage{titlesec}

\newcommand{\myitem}[0]{\scalebox{0.8}{\textbullet}\ }

\usepackage{balance}

\usepackage{enumitem}
\setlist{topsep=0pt, leftmargin=*}

\begin{document}
\pagestyle{empty}

	\title{Query Optimization in the Wild: Realities and Trends}
	
\author{
	\alignauthor Yuanyuan Tian\\
	\affaddr{Gray Systems Lab, Microsoft}\\
	\email{yuanyuantian@microsoft.com}
}
\maketitle
	\begin{abstract}


For nearly half a century, the core design of query optimizers in industrial database systems has remained remarkably stable, relying on foundational principles from System R and the Volcano/Cascades framework. However, the rise of cloud computing, massive data volumes, and unified data platforms has exposed the limitations of this traditional, monolithic architecture. Taking an industrial perspective, this paper reviews the past and present of query optimization in production systems and identifies the challenges they face today. Then this paper highlights three key trends gaining momentum in the industry that promise to address these challenges. First, a tighter feedback loop between query optimization and query execution is being used to improve the robustness of query performance. Second, the scope of optimization is expanding from a single query to entire workloads through the convergence of query optimization and workload optimization. Third, and perhaps most transformatively, the industry is moving from monolithic designs to composable architectures that foster agility and cross-engine collaboration. Together, these trends chart a clear path toward a more dynamic, holistic, and adaptable future for query optimization in practice.


	\end{abstract}
	

\section{Introduction}

Query optimization (QO) is one of the most enduring and formidable challenges in the field of data management. For nearly half a century, since the dawn of relational databases, it has been a cornerstone of academic research, generating a vast body of literature that explores everything from plan enumeration and cost modeling to advanced machine learning based techniques (see surveys~\cite{DBS-077, DBS-082}). Yet, despite decades of brilliant work, query optimization is far from a solved problem. The gap between theoretical possibilities and production reality remains significant, as industrial database systems often favor stability, predictability, and incremental progress over disruptive, high-risk innovation.

If one were to look under the hood of most modern databases today, they would find a design paradigm that has stood the test of time: a cost-based optimizer, built on the foundational frameworks pioneered by System R~\cite{systemR}, Volcano~\cite{volcano}, Cascades~\cite{cascades}, and Starburst~\cite{starburst}. The longevity of this architecture is a powerful testament to its robustness and the foresight of its creators. However, the ground beneath these systems is shifting. The explosion of data volume, the rise of the cloud, the decoupling of storage and compute, and the convergence of once-disparate data engines into unified lakehouse ecosystems present challenges that this classic architecture was never designed to handle. The need for manual tuning, the brittleness of cost models, and the monolithic nature of traditional optimizers are becoming critical bottlenecks.

This paper takes a distinctly industrial perspective, focusing on the past and present of QO as practiced in real-world, production-level database systems. We will examine the challenges these systems face today and argue that the next wave of innovation will not come from replacing the core optimizer, but from augmenting and evolving it in practical, impactful ways. We highlight three promising and pragmatic trends that are already gaining momentum in the industry, each addressing critical limitations of the traditional QO model.

These trends represent a natural evolution, moving from a narrow, static view of optimization to a more dynamic, holistic, and composable one. First, we will explore the collaboration between QO and query execution (QE) to ensure robust performance. Second, we will discuss the convergence of QO with workload optimization (WO), expanding the scope from a single query to optimizing the entire workload. Finally, we will analyze what is perhaps the most transformative shift: the move away from monolithic, customized optimizers toward a more agile and composable QO architecture, enabling faster innovation and cross-engine collaboration in modern data platforms.

\section{QO Evolution: The Past and Present}

\begin{figure*}[tbh]
    \centering
    \vspace{-3mm}
    \includegraphics[width=0.80\textwidth]{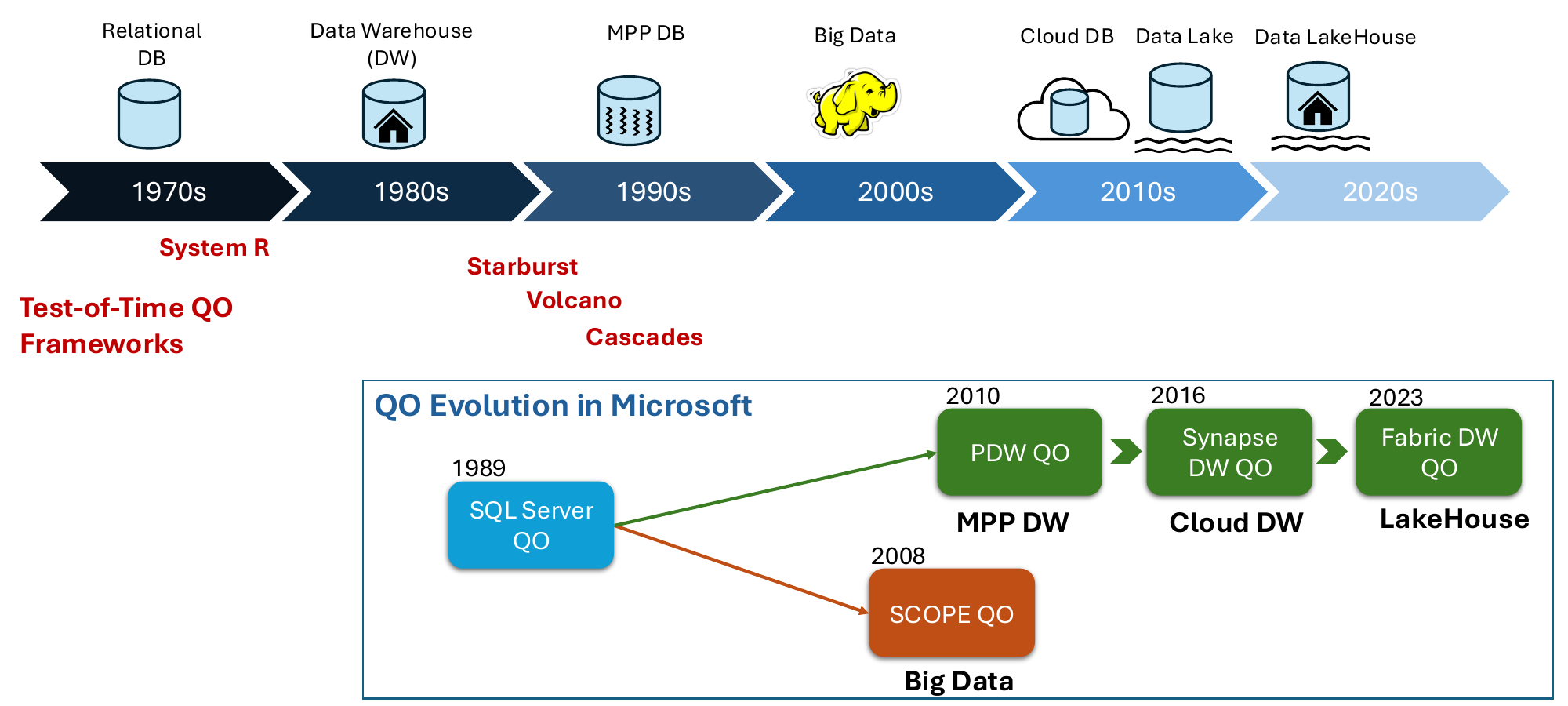}
    \vspace{-3mm}
    \caption{The history of relational databases and QO}
    \vspace{-3mm}
        \label{fig:qohistory}
\end{figure*}

The evolution of QO is best understood through a historical lens. The history of QO is deeply intertwined with the evolution of relational databases. As shown in Figure~\ref{fig:qohistory}, over the past five decades, relational systems have progressed through various architectural shifts, from data warehouses and MPP (Massively Parallel Processing) databases to big data platforms, cloud databases, data lakes, and modern lakehouse architectures. QO has evolved alongside these trends. However, interestingly, if you look closely, many of today’s optimizers still rely on foundational QO frameworks introduced decades ago. Four frameworks in particular, System R~\cite{systemR}, Starburst~\cite{starburst}, Volcano~\cite{volcano}, and Cascades~\cite{cascades} have stood the test of time. System R pioneered the work on query optimization. Starburst’s optimizer remains to be the QO for IBM Db2~\cite{db2} today. Volcano and Cascades introduced extensible and modular architectures that are still widely used in modern database systems~\cite{sqlserver,FabricDW, scope,spanner,lyu2021greenplum,begoli2018apache,soliman2014orca}. The enduring success of these QO architectures is due to their powerful extensibility and adaptability. Although recent years have seen a surge in new relational query engines, their query optimizers are all derived from the same time-tested architectures, primarily the Cascades/Volcano framework, and thus share significant similarities. Below, we outline the standard Cascades/Volcano-based QO architecture adopted by the majority of today's database systems~\cite{sqlserver,FabricDW, scope,spanner,lyu2021greenplum,begoli2018apache,soliman2014orca}.


\noindent \textbf{Reference QO Architecture:} A Cascades/Volcano optimizer generally proceeds through four major phases.

\myitem Parsing/Algebrization: The input SQL query is parsed into an Abstract Syntax Tree (AST) and then converted into an initial logical algebraic tree composed of relational operators such as Join, Project, and Scan.

\myitem Simplification/Normalization: The logical tree is iteratively rewritten into a canonical, semantically equivalent form using transformation rules such as predicate pushdown, join reordering, and projection pruning.

\myitem Cost-Based Exploration: Transformation and implementation rules are applied to explore alternative logical and physical plans, all of which are organized in a MEMO structure that groups logically equivalent expressions; for each group and required physical property, the optimizer recursively computes and memoizes the lowest-cost plan, adding enforcer operators (e.g., Sort, Repartition) when necessary to meet property requirements such as ordering or data distribution.

\myitem Post-Optimization: Additional rewrites and property adjustments are applied to the selected plan, such as adding execution annotations or final physical tweaks, producing the fully optimized execution plan ready for the execution engine.

Implementations of this high-level architecture, however, vary widely. Systems like Spark prioritize rule-based simplification with limited cost-based exploration, while mature engines employ sophisticated cost-based exploration. A common practical strategy in these advanced systems is to break cost-based exploration into sub-stages, each running a full Cascades pass with a subset of rules over a subset of conditions along with a timeout. This strategy helps manage the search space and balance plan quality with compilation time, though the specific stage design highly depends on the workload requirements and system implementation. For instance, SQL Server QO has three stages (OLTP, quick, and full optimization), whereas UQO~\cite{bruno2024unified} (the QO for Fabric DW) roughly divides exploration into two stages: single-node optimization followed by distributed planning.

\subsection{Example: QO Evolution in Microsoft}

To understand the powerful adaptability of the existing QO frameworks, let's use a concrete example to illustrate how an optimizer based on the Cascades framework has evolved to handle new requirements over time (see the lower half of Figure~\ref{fig:qohistory}). 

In 1989, the first version of SQL server~\cite{sqlserver} was shipped with its Cascades-based optimizer. Over the years, Microsoft has successfully evolved the SQL Server QO across multiple database products within the company. 

First, to adapt itself to handle modern analytical workloads, the SQL Server QO was evolved to support column-oriented storage and batch-based query processing~\cite{sqlservercstore}. This was achieved by introducing a new index type called \textit{columnstore index} and a new \textit{Columnstore Index Scan} operator, adding new implementation rules, and defining a new physical property to manage ``row mode" versus ``batch mode", along with with enforcer rules to switch between them.

In 2008, SCOPE~\cite{scope}, an internal big data query engine in Microsoft, forked the SQL Server QO as its base and underwent significant adaptations to suit its new processing environment~\cite{cosmoshistory}. First, to support distributed execution, the optimizer was enhanced with new data exchange operators. It also introduced and reasoned about physical properties like partitioning, sorting, and grouping, using enforcer rules to satisfy operator requirements for these properties~\cite{scopeqo}. Second, to prevent runaway jobs that could consume excessive resources, the team embedded pragmatic safeguards directly into the optimizer. For instance, rules were added to disallow cross products entirely and to limit the results of internal joins to less than 1000x1000 per matching key to prevent cost explosions from small but problematic inputs~\cite{cosmoshistory}.


Simultaneously, Parallel Data Warehouse or PDW~\cite{PDW}, an MPP data warehouse, evolved its optimizer from the SQL Server QO as well. 
To support distributed processing, PDW's optimization pipeline employs multiple optimizers. A front-end optimizer first generates a search space (MEMO) using the standard SQL Server stack. A Distributed Query Optimizer (DQO) then refines this by adding data movement operators, which break the plan into sub-plans, called ``steps". These steps are translated back into SQL and executed by worker nodes, each optimizing its local query fragment independently.


This optimization lineage continued as PDW transitioned into its cloud-based incarnation, Synapse DW~\cite{SynapseDW}, which largely reused the multi-optimizer approach. More recently, as Synapse DW was integrated into the Microsoft Fabric lakehouse ecosystem~\cite{FabricDW}, the optimizer evolved again. The new Unified Query Optimizer (UQO)~\cite{bruno2024unified} refines the design by merging the multiple optimizers back into a single, cost-based optimizer for generating distributed plans. It achieves this by incorporating distribution-related physical properties directly into the MEMO, along with implementation and enforcer rules that manage data distribution requirements for operators.

\section{Is QO a ``Solve'' Problem?}

\begin{table*}
    \centering
    \caption{The assumptions for QO vs the reality}
    \vspace{-1em}
    \renewcommand{\arraystretch}{1.5}
    \scriptsize
    \begin{tabular}{|c|c|c|} \hline
         & \textbf{Assumption} & \textbf{Reality} \\ \hline
      \textbf{Workload}   & One query at a time & Queries run simultaneously\\ \hline
       \textbf{Additivity}  & Costs can be simply added together & Operations interleave\\ \hline
       \textbf{Independence}  & Predicates are independent & Predicates are often correlated\\ \hline
       \textbf{Subsumption}  & Joins are PK-FK joins  & Many-to-many joins\\ \hline
       \textbf{Weighting}  & Certain types of cost (I/Os, memory, CPU) dominate & Hardware \& architecture have changed dramatically\\ \hline
       \textbf{Model Detail}  & Detailed models are more accurate & More details add more assumptions!\\ \hline
    \end{tabular}
    \label{tab:qoassumption}
\end{table*}


If today's databases still use QO frameworks from decades ago, does that mean QO is a ``solved" problem? Guy Lohman, the QO expert behind Starburst, addressed this very question in his 2014 ACM blog post~\cite{Lohmanblog} and again in his 2017 talk at BTW~\cite{Lohmanbtw}, concluding that the answer is no. 

The core issue lies in the cost model, which serves as the optimizer's foundation but is built on a series of outdated assumptions (as detailed in Table~\ref{tab:qoassumption}). It assumes that queries are executed one at a time, the costs of different operators can be simply aggregated, predicates are independent of each other, all joins are primary key-foreign key joins, certain types of costs, such as I/O, dominate others, and that a more detailed model enhances accuracy. However, in reality, queries run concurrently, operations interleave, predicates are often correlated, there are many-to-many joins, hardware and database architecture have undergone significant changes over the years, and increased detail introduces more assumptions into the cost model, ultimately making the QO even more brittle. The shift to the cloud has particularly stressed traditional QO. Cost models must now incorporate the complexities of multi-tier storage (SSD, HDD, cloud object stores) and network costs, making them even more brittle. 


This view is strongly supported by empirical evidence. A study by Leis et al.~\cite{LeisGMBK015} empirically evaluated the main components of a classic QO including cardinality estimation, the cost model, and plan enumeration techniques, and came to the same conclusion. This study found that all tested database systems routinely produce large errors in their cardinality estimates. The join size estimators based on the independence assumption are particularly fragile. In addition, the study found that simple cost models were found to be sufficient, as their errors were dwarfed by the errors in cardinality estimation.



\section{Practical Trends in QO}
This section will discuss three trends in the industry aimed at addressing some of the QO problems in practice. 

\subsection{Collaboration between QO and QE}
Given that query optimization can often be error-prone, the goal of the QO has shifted, from always trying to find the absolute best plan to primarily avoiding really bad ones. As a result, QO and QE now need to work hand-in-hand to ensure consistently good performance. 

One common strategy is to build robust query execution engines that are more resilient to suboptimal plans. For example, most modern databases employ techniques like Bloom filters or bitmap filters to pre-filter tables and reduce the size of join inputs~\cite{db2,sqlserver,oracle,spark}. This helps the execution engine tolerate poor join orderings decided by the QO and still deliver acceptable performance. In particular, the authors of~\cite{sqlbf} proved that the synergistic interaction between SQL Server's bitmap filters, its pull-based execution model, and the Cascades optimizer allows the system to generate \textit{instance-optimal} plans for acyclic joins, thereby achieving results equivalent to those of Yannakakis's algorithm.

\begin{figure}[tbh]
    \centering
    \vspace{-4mm}
    \includegraphics[width=0.2\textwidth]{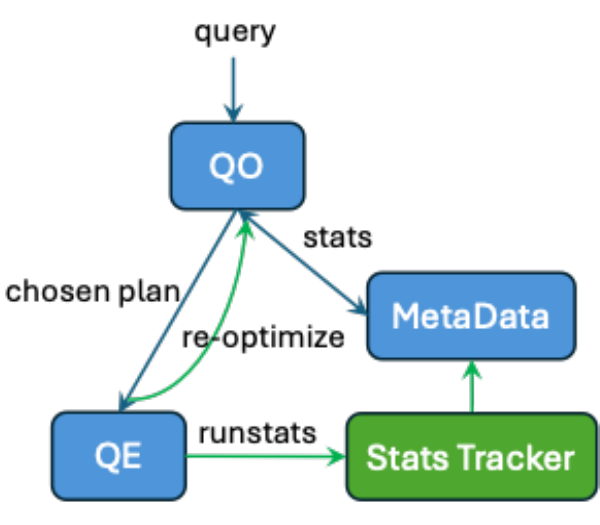}
    \vspace{-3mm}
    \caption{The adaptive QO architecture}
    \vspace{-3mm}
        \label{fig:adaptiveqo}
\end{figure}

Another approach where QO and QE collaborate to ensure query performance is adaptive query optimization. In this strategy (see Figure~\ref{fig:adaptiveqo}), the QE gathers runtime statistics and based on this feedback, can trigger re-optimization to adjust the execution plan dynamically. 

Many modern databases implement some form of adaptive optimization. For instance, the adaptive join feature of SQL server~\cite{adjoinsqlserver} and Oracle~\cite{oracleqo} can switch from a hash join to a nested loops join at runtime, depending on observed cardinalities. Google BigQuery defaults to a shuffle join but may switch to a broadcast join during execution if the data volume allows~\cite{dremel}. More recently, Databricks introduced adaptive query execution~\cite{adodatabricks}, which uses runtime statistics from completed stages to choose between shuffle and broadcast joins, adjust the degree of parallelism, and restart scans with dynamically generated Bloom filters.

\subsection{Converging QO with WO}

\begin{figure}[tbh]
    \centering
    \vspace{-3mm}
    \includegraphics[width=0.25\textwidth]{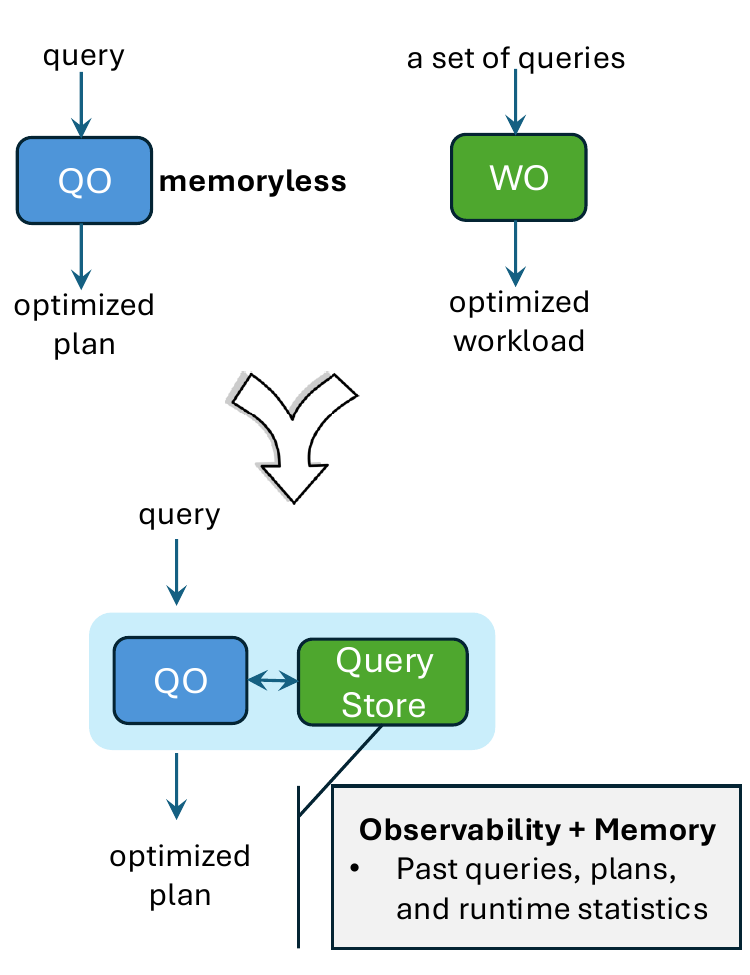}
    \vspace{-3mm}
    \caption{The convergence of QO and WO}
    \vspace{-3mm}
        \label{fig:qowo}
\end{figure}

The second trend is the convergence of QO and WO. Traditional QO is memoryless. It optimizes each query without considering past executions. On the other hand, WO takes in query traces and optimize for the entire workload, such as creating indexes or materialized views. Modern database systems are converging QO and WO by utilizing workload-level insights to optimize individual queries more effectively (see Figure~\ref{fig:qowo}). 

For example, SQL Server uses the Query Store to persist historical execution plans for a query and monitors query performance to detect plan changes or regressions over time~\cite{querystore}. It can also apply Automatic Plan Correction to fall back to a last known best plan if regression is detected~\cite{plancrct}. Similarly, Oracle's SQL Plan Management feature~\cite{sqlplanmgmt} maintains a set of valid plans for each query. The optimizer can then choose from these baseline plans, usually the one with the lowest cost.

\subsubsection{About Learned QO}
In recent years, there has been a significant increase in the application of Machine Learning (ML) techniques to enhance query optimization, a research area now commonly referred to as ``learned QO"~\cite{neo, bao, skinnerdb, fastgres, cedata, Kipf2018LearnedCE, hybridCE}.
The core premise behind all learned QO approaches is that queries are often recurrent and similar, enabling these methods to  learn from the past to improve the future. So, essentially all learned QO approaches are embodiments of the convergence of QO and WO. 

Despite the rapidly growing body of research, the industry remains hesitant to deploy learned QO techniques in real-world production systems. As discussed in~\cite{autods}, this gap between academia and industry is rooted in several key factors: production systems are far more complex than research prototypes, and they have strict requirements for explainability and debuggability that many ML models cannot meet; furthermore, there is a significant risk of performance regressions when models trained on historical data encounter evolving workloads, and the high cost of training and inference remains a major barrier.

Nevertheless, there are some industrial efforts in applying the learned QO approaches to production systems~\cite{cefeedback, dopfeedback}. \textit{In general, the ``pacemaker approach" of using ML to enhance existing QO is often more feasible than the ``heart transplant approach" of replacing the QO entirely with a learned model.} 

Let's illustrate examples of pacemaker approaches. SQL Server has a number of Intelligent Query Processing (IQP) features~\cite{iqp} where the system learns from its mistakes during runtime and automatically applies corrective measures to improve future performance. For example, the cardinality estimation (CE) Feedback feature~\cite{cefeedback} observes runtime differences between estimated and actual row counts and automatically adjusts the optimizer's CE model assumptions (fully independent vs. partially correlated vs. fully correlated) for future executions of repeating queries. In a similar vein, its Degree of Parallelism (DoP) Feedback feature learns and enforces the optimal DoP for repeating queries~\cite{dopfeedback}. A recent work~\cite{qoaas} improved the SQL Server cost model by using ML to automatically tune the constant parameters in the cost model offline for a target workload. Adapting from techniques in Bao~\cite{bao}, QOAdvisor~\cite{knobadvisor1, knobadvisor} recommends rule hints for Microsoft Scope QO. It had to make practical adjustments to manage the system complexity, like incremental plan changes, cost-efficient experimentation with a contextual bandit model, and a validation model to prevent regressions. Google BigQuery recently introduced History-Based Optimizations~\cite{bigquery-hbo}, which learn patterns and characteristics from prior query executions to generate efficient execution plans. Key optimizations include join pushdown, semi-join reduction, build/probe side swapping for hash joins, and dynamic adjustment of DoP. Redshift leverages its Automated Materialized Views (AutoMV) feature~\cite{redshift-automv} to boost query performance, employing ML and workload monitoring to automate the creation and management of views. Databricks’ recent Predictive Optimization feature~\cite{databricks-po} uses ML to collect and maintain table statistics in a smart way.

These above methods share a common strategy: they preserve the existing QO with minimal modification, while using ML to find the optimal settings and ``knobs" for it to operate under. Furthermore, a successful learned QO implementation typically relies on two critical components: comprehensive history tracking (e.g., query logs, runtime statistics) and robust feedback mechanisms to continuously refine performance.

\subsubsection{Related Directions Beyond Learned QO}
While not intrinsic to QO, \textbf{ML-driven storage optimization} is increasingly used to improve performance. Systems like Databricks (Predictive I/O~\cite{databricks-predictive-io}, Liquid Clustering\cite{databricks-liquid}) and Amazon Redshift~\cite{redshift-auto-tuning} use ML to automate physical layout decisions, such as clustering keys, sort orders, and data distribution, creating data organizations that the QO can exploit for faster execution. Regarding \textbf{Generative AI}, while LLMs have revolutionized user interfaces (e.g., text-to-SQL), their adoption in core QO remains experimental at best. The strict requirements for determinism, low latency, and absolute correctness in plan generation currently conflict with the probabilistic nature and high inference costs of LLMs, preventing their use in the critical path of production QO.

\subsection{From Customized to Composable QO}

The third major trend is the transition from customized to composable QO. This transition is primarily influenced by two significant, industry-wide developments.

  \noindent  \textbf{The Convergence of Data Platforms}: The first driver is the industry's shift away from siloed data engines toward highly integrated and converged solutions. Microsoft Fabric~\cite{fabric} is a notable example, creating a unified lakehouse where different compute engines operate on the same copy of data in OneLake stored as Delta Parquet tables. Within this ecosystem, engines like Fabric Spark and Fabric DW share the same physical data, have similar architectures and usage patterns, yet differ significantly in optimizer maturity, with DW's QO being far more advanced. 
  This specific combination makes a shared QO far more valuable than in a single engine system where decoupling the QO offers lower ROI.
  
    
  \noindent  \textbf{The Rise of Composable Database Architectures}: Concurrently, there has been a fundamental move from monolithic to composable database designs~\cite{composabledb}, spurred by cloud adoption and the decoupling of storage and compute. This modularity is further enabled by the widespread adoption of open standards such as Parquet~\cite{parquet}, Arrow~\cite{arrow}, and Substrait~\cite{substrait}. In the Lakehouse ecosystem, this trend is evident in the proliferation of open table formats (Delta Lake\cite{delta-lake}, Hudi\cite{apache-hudi}, and Iceberg\cite{apache-iceberg}) and interoperability layers like XTable\cite{xtable} and UniForm\cite{uniform}, alongside unified catalog initiatives such as OpenHouse\cite{openhouse}, Polaris\cite{apachepolaris}, and Unity Catalog\cite{unity}.   
  In addition, there has been a surge in open-source projects focused on reusable database components. On the QE side, projects like Velox~\cite{pedreira2022velox} and Datafusion~\cite{datafusion} exemplify this, while on the QO side, general-purpose libraries like Calcite~\cite{calcite} and Orca~\cite{orca} are being developed to unify and share core QO components. 

Both Orca and Calcite represent ``QO-as-a-Library" design philosophy, where a modular, cost-based optimizer can be embedded within and customized for different host systems. Built upon the Volcano/Cascades paradigm, they each provide a foundation for adding custom rewrite rules and operators. While Orca and Calcite may not directly solve today's core optimizer challenges, their modular and extensible architectures are crucial enablers. By fostering faster innovation, improving engineering efficiency, and reducing time-to-market for new engines, they create the foundation needed to develop and deploy the next generation of QO technologies that can address these modern problems.


Recent work~\cite{qoaas} further elevates composability with the proposal of ``QO-as-a-Service" (QOaaS) for unified ecosystems like Microsoft Fabric. By externalizing the QO into an independent service that communicates with engines via RPC, QOaaS retains the benefits of QO libraries while enabling new capabilities like independent scaling, workload-aware tuning, and cross-engine optimization. The architecture ensures interoperability by using a standard plan specification like Substrait~\cite{substrait} and establishes a closed-loop system. To achieve this, the QOaaS proposal introduces three key components: a Query Insight Store to capture historical plans and runtime statistics; an External Tuner Plugin to enable data-driven tuning by external processes; and a Config/Action Store to feed optimized configurations back into the core optimizer for continuous improvement.

\section{Conclusion and Discussion}

In conclusion, the foundational architecture of QO in modern databases remains largely unchanged from the frameworks proposed five decades ago. Yet, as we have discussed, QO is far from a solved problem. The way forward is not to endlessly polish the existing model, but to address its foundational limitations with practical, forward-looking solutions. This paper has highlighted three pragmatic trends gaining momentum in the industry: the collaboration between QO and QE; the convergence of QO with WO; and the transformative shift from monolithic to composable QO architectures. 

While feedback loops and workload-level optimization inherently introduce telemetry and storage overheads, production systems mitigate these costs through adaptive sampling (logging only identifying/expensive queries) and asynchronous processing (performing heavy analysis off the critical path). Furthermore, this ``tax" is viewed as an essential insurance policy; the cost of storing historical statistics is negligible compared to the resource waste and SLA violations caused by a single catastrophic regression. In modern cloud architectures, this trade-off is increasingly favorable, prioritizing system resilience over raw minimal overhead.

 Finally, as noted in~\cite{stillasking}, a major challenge for the progression of an industry QO is performance regression. No matter how well a method improves average workload performance, a small number of slower queries can be a deal-breaker for adoption. To varying degrees, the three trends highlighted in this paper serve as an antidote to this stagnation. QO and QE collaboration acts as the airbag, resilient enough to handle, and even correcting, QO mistakes at runtime. By merging QO and WO, we move from stateless guessing to stateful learning, allowing the system to remember what works and avoid repeating mistakes (but falling back to safe defaults for unobserved queries). A composable QO service, empowered by a Query Insight Store and pluggable interfaces, creates the infrastructure necessary to test and integrate innovations safely, making the progression of QO both safer and faster.

\bibliographystyle{abbrv}
\small
\bibliography{reference}

\end{document}